\documentclass[pdflatex,sn-basic]{sn-jnl}

\jyear{2021}%

\newcommand{\bfit}[1]{\textbf{\textit{#1}}}

\begin{document}

\title[Dust dynamics in planet-forming discs in binary systems]{Dust dynamics in planet-forming discs in binary systems}

\author*[1]{\fnm{F.} \sur{Zagaria}}\email{fz258@cam.ac.uk}
\author[2,3,4]{\fnm{G. P.} \sur{Rosotti}}
\author[3]{\fnm{R. D.} \sur{Alexander}}
\author[1]{\fnm{C. J.} \sur{Clarke}}
\affil[1]{\orgdiv{Institute of Astronomy}, \orgname{University of Cambridge}, \orgaddress{\street{Madingley Rd}, \city{Cambridge}, \postcode{CB3 0HA}, \country{UK}}}
\affil[2]{\orgdiv{Leiden Observatory}, \orgname{Leiden University}, \orgaddress{\street{Niels Bohrweg 2}, \city{NL-2300 RA Leiden}, \postcode{PO Box 9513}, \country{the Netherlands}}}
\affil[3]{\orgdiv{School of Physics and Astronomy}, \orgname{University of Leicester}, \orgaddress{\street{University Rd}, \city{Leicester}, \postcode{LE1 7RH}, \country{UK}}}
\affil[4]{\orgdiv{Dipartimento di Fisica}, \orgname{Universit\`a degli Studi di Milano}, \orgaddress{\street{Via Giovanni Celoria 16}, \city{Milano} \postcode{I-20133}, \country{Italy}}}
  

\abstract{In multiple stellar systems, interactions among the companion stars and their discs affect planet formation. In the circumstellar case, tidal truncation makes protoplanetary discs smaller, fainter and less long-lived than those evolving in isolation, thereby reducing the amount of material (gas and dust) available to assemble planetary embryos. On the contrary, in the circumbinary case the reduced accretion can increase the disc lifetime, with beneficial effects on planet formation. In this chapter we review the main observational results on discs in multiple stellar systems and discuss their possible explanations, focusing on recent numerical simulations, mainly dealing with dust dynamics and disc evolution. Finally, some open issues and future research directions are examined.}




\maketitle

\section{Introduction}\label{sec:1}
Multiplicity is common among main sequence stars \citep{DuquennoyMayor1991,Raghavan2010,DucheneKraus2013,Tokovinin2014,MoeDiStefano2017}, and even more around young stellar objects \citep{Chen2013}. The increasing census of exoplanets detected in binaries \citep{Hatzes2016,Martin2018} indicates that planet formation is viable in multiple stellar systems. Two orbital configurations are allowed \citep{Dvorak1986}: (a) circumstellar or ``S(atellite)-type'' planets have a semi-major axis smaller than the binary one, and orbit one of the two binary components \cite[e.g.,][]{Hatzes2003,Dumusque2012}; (b) circumbinary or ``P(lanet)-type'' planets have a semi-major axis larger than the binary one, and orbit both binary components \cite[e.g.,][]{Doyle2011,Welsh2012}. 

Radial velocity and high-resolution imaging exoplanet surveys established that multiplicity has a substantial impact on the occurrence of circumstellar planets \cite[e.g.,][]{Wang2014,Wang2015_b,Wang2015_c,Ngo2016,Kraus2016,Matson2018,Ziegler2020}. A combination of these datasets confirmed \cite[see][]{MoeKratter2021} that the occurrence of S-type planets is progressively reduced in binaries closer than $a\approx200\ {\rm au}$. In particular, the stellar companion fraction among planet hosts is as small as $\approx15\%$ of the field for separations $a\approx10\ {\rm au}$, and is completely suppressed for $a\lesssim1\ {\rm au}$. Those estimates can be improved only if future studies will be able to target control samples of field stars similar to planet hosts, in order to reduce biases and systematic uncertainties \citep{MoeKratter2021}.

Instead, despite the recent improvements with \textit{TESS} data \citep{Kostov2020,Kostov2021}, the statistics of circumbinary planets remain limited and are more debated. In close \textit{Kepler} eclipsing binaries, the occurrence rate of giant planets coplanar with their binary host is $\approx10\%$ \citep{MartinTriaud2014,Armstrong2014}. This result is consistent with the single-star one, but can hint at a higher frequency of circumbinary planets, if systems with higher mutual inclinations are present yet not detected. Alternatively, a primordial alignment of the planet and binary orbit within $\lesssim3^\circ$ can be deduced, when selection biases are considered \citep{Li2016}.

The recently-issued \textit{Protostars and Planets VII} chapter on multiplicity \citep{Offner2022_PPVII} thoroughly reviewed the connections between stellar binarity and planet-formation. In this paper, instead, we focus on the intermediate T~Tauri phase, to highlight the effects of binarity on protoplanetary disc evolution, and how binarity affects the reservoir of planet-forming material on secular timescales. Firstly, in \autoref{sec:2} we summarise the observational results on discs in binary systems. Then in \autoref{sec:3}, \ref{sec:4} and \ref{sec:5} we present the solutions proposed by recent numerical simulations, with particular focus on the case of dust. We also display the results of new simplified models of dusty circumbinary discs. Finally, in \autoref{sec:6} we discuss some open questions and possible future directions, and in \autoref{sec:7} we draw our conclusions.

\section{Observations of discs in multiple systems}\label{sec:2}
On observational grounds, there is general consensus on stellar multiplicity substantially influencing disc evolution. Infrared excess signatures showed that the fraction of stars with discs decreases in multiple systems with decreasing binary separation \citep{Cieza2009,Kraus2012,Cheetham2015}: in binaries closer than $a=40$~au, after 1 to 2 Myr, stars are two to three times less likely to host a disc than singles. Similar depletion factors are expected over the next Myr, suggesting that multiplicity affects especially young discs \citep{Kraus2012,Barenfeld2019}. Using accretion signatures, analogous results were obtained \citep{Daemgen2012,Daemgen2013,Daemgen2016}: in Chamaeleon~I (age 2 to 3 Myr) the fraction of Br$\gamma$ emitters among $a<100$~au binaries is six time smaller than in wider binaries and singles \citep{Daemgen2016}.

Unresolved observation of discs in binary systems in the (sub-)mm, using single-dish telescopes, showed that multiplicity also affects continuum luminosities, with discs being fainter in closer binaries \citep{Jensen1994,Jensen1996,OsterlohBeckwith1995}. The advent of interferometry made it easier to resolve these systems and detect discs, first around primaries and in small samples \citep{JensenAkeson2003,Patience2003}, then for several binaries also around secondaries \citep{Harris2012}. In Taurus (age 1 to 2 Myr) these data suggested that primary discs are generally brighter than secondary, and supported the evidence for an increasing trend of continuum emission with binary separation: discs in pairs wider than $a\approx300$~au are as luminous as singles, those with separations between 30 and 300~au have five time less flux, and those closer than $a\approx30$~au are fainter by an additional factor of five\footnote{On the contrary, in the older Upper Sco star-forming region (age 5 to 11 Myr), the distributions of fluxes in close binaries and singles are indistinguishable \citep{Barenfeld2019}.}. Instead, no clear relation between continuum flux and binary mass ratio was found \citep{Harris2012}.

Recent ALMA surveys confirmed these results \citep{AkesonJensen2014,Cox2017,Akeson2019,Manara2019,Zurlo2020,Zurlo2021,Panic2021}. However, the higher disc detection fractions allowed to show that discs around primaries are more massive than those around secondaries only because of their different stellar masses: primaries and secondaries have similar luminosity distributions, both incompatible with that of singles, when normalised by the stellar mass \citep{Akeson2019}.

The unprecedented angular resolution provided by ALMA allowed to resolve circumstellar discs in binary systems, showing that dust emission is more compact \citep{Cox2017,Manara2019,Zurlo2020,Zurlo2021} and has a steeper outer edge \citep{Manara2019} in binaries than in singles. Also, primaries have larger discs than secondaries \citep{Manara2019}. As in singles, gas sizes are larger than dust sizes, establishing that the latter do not trace the truncation radius \citep{Rodriguez2018,Rota2022}. Multiplicity also affects the ratio between gas and dust sizes: when the radius enclosing 95\% of the disc flux is considered, the $R_{\rm 95,CO}/R_{\rm 95,dust}$ ratio is larger in binaries (on average $\approx4.2$, \citealt{Rota2022}) than in singles (on average $\approx2.8$, \citealt{Sanchis2021,Long2022}). This ratio is also larger for secondary discs (on average $\approx5.0$) than for primaries (on average $\approx3.9$), and does not correlate with separation \citep{Rota2022}. Finally, gas kinematics allowed to study the interaction between a disc and an external companion as in the case of the candidate fly-bys RW~Aur \citep{Dai2015} or UX~Tau \citep{Menard2020}.

As for accretion signatures, disc surveys in multiple systems are rare because of the high spectral and angular resolution required. In fact, disentangling the accretion signatures of each component remains challenging \cite[e.g.,][]{Frasca2021}. HST and IRTF data showed that the accretion rates of single and $a\approx10$ to 100~au binary stars in Taurus are similar. In addition, primaries accrete more than secondaries, particularly in systems with small mass ratios, as is expected from their different stellar masses \citep{WhiteGhez2001}. These results were later confirmed in the ONC and Chamaeleon~I  using VLT/NACO \citep{Daemgen2012,Daemgen2013}. Recent results of multiple-epoch observations showed accretion variability from 0.2 to 0.5~dex on the time scale of some days to four years \citep{Fiorellino2022b,Zsidi2022b}. When available, dust masses and accretion rates were used to estimate the disc accretion timescale $t_{\rm acc}=M_{\rm disc}/\dot{M}_{\rm acc}$, which is often interpreted as a measure of the disc age. However, results suggest that in binaries $t_{\rm acc}$ is shorter than in singles, and much smaller than the disc age inferred from stellar evolutionary tracks \citep{WhiteGhez2001}. 

The recent advent of new instruments like ALMA and the Spectro-Polarimetric High-contrast Exoplanet REsearch (SPHERE) at VLT has also been revolutionising our understanding of circumbinary discs. Despite being not as ubiquitous as S-type ones \citep{Czekala2019}, there are notable examples of (sub-)mm and near-infrared polarimetric observations of P-type discs as well. Common features in these systems are dust cavities surrounded by ring(s), either circular and azimuthally symmetric \citep[e.g.,][]{Kurtovic2018,Martinez-Brunner2022} or, more often, eccentric (as in the case of HD~142527, where $e\approx0.32$ to 0.44, \citealt{Garg2021}) and showing clear asymmetries in the dust and/or CO isotopologues \citep{Dutrey2014,Phuong2020,Kraus2020,Long2021,Ragusa2021,Kurtovic2022}. In the latter case, spirals \citep{Keppler2020,Long2021,Garg2021}, warps \citep[e.g.,][]{Kraus2020,Long2021}, stream-like filaments \citep{Phuong2020,Keppler2020} and shadows cast by misaligned inner disc(s) \citep{Ginski2018,Keppler2020,Kraus2020,Martinez-Brunner2022} are also detected. These features are generally interpreted as due to binary-disc interactions. In addition to disc morphology, ALMA observations are useful to provide tight constraints on binary masses and orbital parameters through gas kinematics \citep{Czekala2015,Czekala2016,Czekala2017,Czekala2021}. 

Nonetheless, circumbinary disc statistics are still limited, and it remains unclear whether or not their lifetimes are significantly different from single-star systems \citep{Kuruwita2018,Kounkel2019}. Indeed, there may be a dichotomy in outcomes: disc formation is usually suppressed in binary systems, but some of the longest-lived and most massive known discs turn out to be circumbinary (e.g., V4046~Sgr, with an age of $\approx20$~Myr, \citealt{Rosenfeld2013,Martinez-Brunner2022}). A larger census of circumbinary discs is needed to address this question.

\section{Theory of disc truncation}\label{sec:3}
The interaction between a gaseous disc and an embedded satellite (e.g., a planet or a stellar companion) has been investigated since the late '70s, either tidally \citep{PapaloizouPringle1977}, in the so-called ``impulse approximation" \citep{LinPapaloizou1979a,LinPapaloizou1979b}, or assuming that satellite-disc interactions are excited at specific locations in the disc \citep{Lynden-BellKalnajs1972,GoldreichTremaine1979,GoldreichTremaine1980,Meyer-VernetSicardy1987}. Both approaches lead to the conclusions that: (a) the primary disc exerts a torque on the secondary, which causes the satellite to migrate, and (b) as a back-reaction, the secondary exerts a torque on the primary disc, opening a gap around the satellite, which ultimately leads to disc truncation. As a consequence, a binary can host up to three discs, two circumstellar (orbiting each star separately) and a circumbinary (orbiting both stars).

The location where a disc is truncated depends on several parameters. In the case of circular coplanar tidally-interacting binaries, truncation occurs when the viscous and tidal torques balance. Because both these terms scale linearly with viscosity, the truncation radius does not depend on viscosity \citep{PapaloizouPringle1977}. This is true also when the balance of resonant and viscous torques are considered, giving similar estimates for the truncation radius \citep{ArtymowiczLubow1994}. Instead, when the binary is eccentric, the disc viscosity is also important in determining where a disc is truncated \citep{ArtymowiczLubow1994}. Recently, \cite{Manara2019} proposed the following analytical expression for the truncation radius:
\begin{equation}
    R_{\rm t}=R_{\rm Egg}\left(\alpha e^\beta+\gamma\mu^\delta\right),
\end{equation}
where $R_{\rm Egg}$ is an estimate for the Roche lobe radius \citep{Eggleton1983}, the minimum distance from the star where the satellite would be ripped apart by tidal forces, $e$ is the binary eccentricity, $\mu=m_2/(m_1+m_2)$, and $\alpha$, $\beta$, $\gamma$ and $\delta$ are free parameters chosen to reproduce the results of \cite{ArtymowiczLubow1994}. $\gamma=0.88$ and $\delta=0.01$ are determined by fitting the estimates for truncation radius of \cite{PapaloizouPringle1977} in circular binaries, while $\alpha$ and $\beta$ depend on the Reynolds number and are tabulated in \cite{Manara2019} for different values of $\mu$. Finally, considerations based on the stability of test particle orbits showed that a disc can be truncated where particle streamlines intersect \citep{Paczynski1977,RudakPaczynski1981,HolmanWiegert1999,Pichardo2005}. This approach leads to similar estimates of the truncation radius. Most recently, the resonant theory was extended to include the effects of disc inclination with respect to the binary orbital plane \citep{MirandaLai2015}.

\section{Circumstellar disc simulations}\label{sec:4}

\subsection{Can disc models explain the observed binary \textit{planets}?}\label{sec:4.1}
Numerical studies of circumstellar protoplanetary discs in binary systems mainly focused on planetesimal evolution. Earlier phases were almost only considered to analyse the conditions for planet formation in very close ($a\approx20$ to 30 au), highly viscous ($\alpha\lesssim10^{-2}$) binaries. Here we briefly summarise these results and refer to \cite{ThebaultHaghighipour2015,MarzariThebault2019} for their connections to planet formation.

\cite{Jang-Condell2008,Jang-Condell2015} explored the range of disc and stellar parameters compatible with core accretion, concluding that $\gamma$-Cephei \citep{Hatzes2003}, and other observed binary systems hosting planets, may have had discs massive just enough to assemble them. However, their models did not consider gas and dust evolution. There is some debate regarding how eccentric we should expect circumstellar discs to be, because eccentricity is potentially detrimental to planet growth because of the high relative velocities between planetesimals. \cite{KleyNelson2008} first studied the dynamical effect of a close binary companion on a gaseous disc, showing that spiral structures and eccentric modes can be triggered at pericenter (even though these results are sensitive to the inner boundary conditions, \citealt{Kley2008}). Later \cite{MullerKley2012} and \cite{Marzari2009,Marzari2012} argued that eccentricity was overestimated, and a proper treatment of viscous heating and radiative cooling, as well as the inclusion of disc self-gravity, imply lower disc eccentricities. However, most recently \cite{Jordan2021} showed that these works may have suffered from too low numerical resolution, and that more realistic viscosities lead to $e\approx0.2$, leaving this problem open.

As for dust, simulations of growth and transport have traditionally been challenging also in the case of single-star discs \citep{Brauer2008,Birnstiel2010}. Based on their 2/3D gas simulations, \cite{Nelson2000,PicognaMarzari2013} proposed that, in binaries of separation $a\approx30$ to 50~au, the strong disc heating, due to spirals and mass transfer between the discs, sublimates volatiles and inhibits dust coagulation. \cite{Zsom2011} considered a similar problem coupling a simplified growth/erosion and dust drift model to gas simulations. They showed that the presence of a binary companion reduces the mass and stopping time of the solids, particularly if the disc is eccentric. Even though grains can grow between spirals, \cite{Zsom2011} argued that the lack of a reservoir of dust in the outer disc makes radial drift more efficient in binaries. 

To sum up, these works paint a consistent picture that planet formation in binaries faces more challenges than in single systems. The observational evidence that planets in binary systems nevertheless exist implies that planet formation must then be a fast process, unless viscosity is low enough \citep{MullerKley2012} and dust drift is slowed down \citep{Zsom2011}.

\subsection{Can disc models explain the observed binary \textit{discs}?}\label{sec:4.2}
A population-oriented approach has also been sought, in order to systematically compare protoplanetary disc simulations and observations. In these works, circular coplanar binary discs are considered and tidal truncation is modelled impeding any exchange of material at the expected truncation radius.

\paragraph{Gas: explaining the relative binary disc fraction} 
Evidence of systems where only one component shows IR excess \citep{MoninPPV,ReipurthPPVI} motivated the study of the relative lifetime of primary and secondary discs. 

When discs are evolving viscously, the disc-clearing timescale is set by the truncation radius, because of the suppression of viscous spreading. So, for $q\ll1$, the secondary disc is expected to clear first, because it is truncated at a smaller radius than the primary. Instead, for $q\gtrsim0.5$, primary and secondary discs have similar sizes and their relative lifetime depends on the initial conditions \citep{Armitage1999}. However, X-ray photoevaporation can change this picture. While single-star discs are cleared inside-out, in binary discs dispersal takes place inside-out or outside-in depending on the relative location of the gravitational radius, $R_{\rm g}=Gm_i/c_{\rm s}^2$ with $i\in\{1,2\}$, and truncation radius, $R_{\rm t}$. In close binaries, where the disc disappears before photoevaporation becomes significant, secondaries live less long than primaries, because their truncation radius is smaller. Instead, in wide binaries, where photoevaporation can open up a hole, the primary lives less long, because its photoevaporation rate is higher \citep{RosottiClarke2018}. This different qualitative behaviour also led to the prediction of fewer transition discs in binaries than in singles \citep{RosottiClarke2018}.

When simulated populations of discs, with observationally-motivated initial conditions, were compared with data, a good agreement was found, both with the disc fraction of Taurus binaries as a function of the companion separation, and with the relative lifetime of primary and secondary discs as a function of the binary separation and mass ratio \citep{RosottiClarke2018}. However, the small samples and some observational biases made the comparison more challenging.

\begin{figure}[t!]
    \centering
    \includegraphics[width=\textwidth]{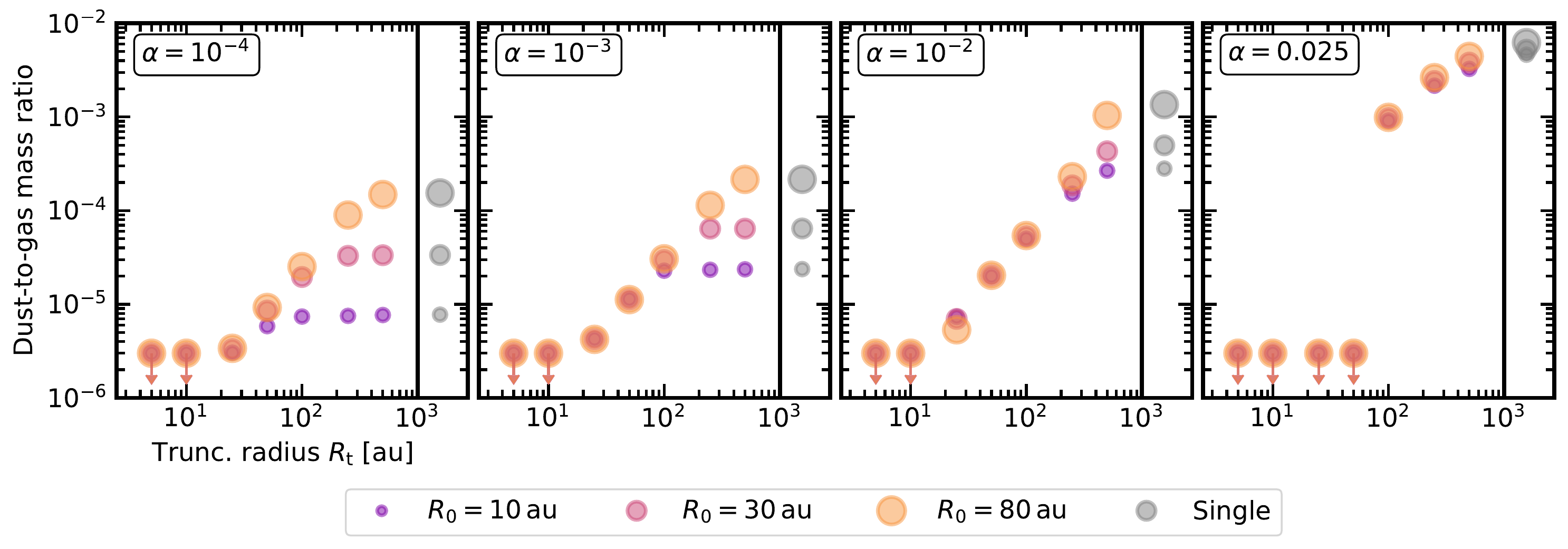}
    \caption{The dust-to-gas mass ratio in circumstellar binary discs increases with the companion separation (2~Myr snapshot). Single discs are in grey while different colours/sizes display different values of $R_0$.}
    \label{fig:1}
\end{figure}

\paragraph{Dust: explaining dust disc sizes}
To explore the effect of binarity on dust dynamics, \cite{Zagaria2021_theory} coupled a simplified prescription for dust growth/fragmentation \citep{Birnstiel2012} and dust transport \citep{LaibePrice2014,Booth2017} to viscous binary evolution. Their main result is displayed in Fig.~\ref{fig:1}, where the dust-to-gas mass ratio is plotted as a function of the truncation radius, after 2~Myr, for different values of $\alpha$. The figure shows that the smaller the truncation radius is, the faster binary discs lose dust: drift is more efficient because truncation progressively reduces the reservoir of small grains in the outer disc. Two different disc-clearing mechanisms can be seen: for a small $\alpha$, discs lose dust fast because of drift, but are longer-lived in the gas (as suggested by previous works and supported by data, see \citealt{Barenfeld2019}); for a large $\alpha$, discs are initially fragmentation-dominated and retain more dust, but are dispersed faster, because of their shorter lifetime. Finally, we note that the correlation intriguingly resembles the behaviour of the companion fraction suppression in planet hosts of \cite{MoeKratter2021}.

\begin{figure}[t!]
    \includegraphics[width=0.5\textwidth]{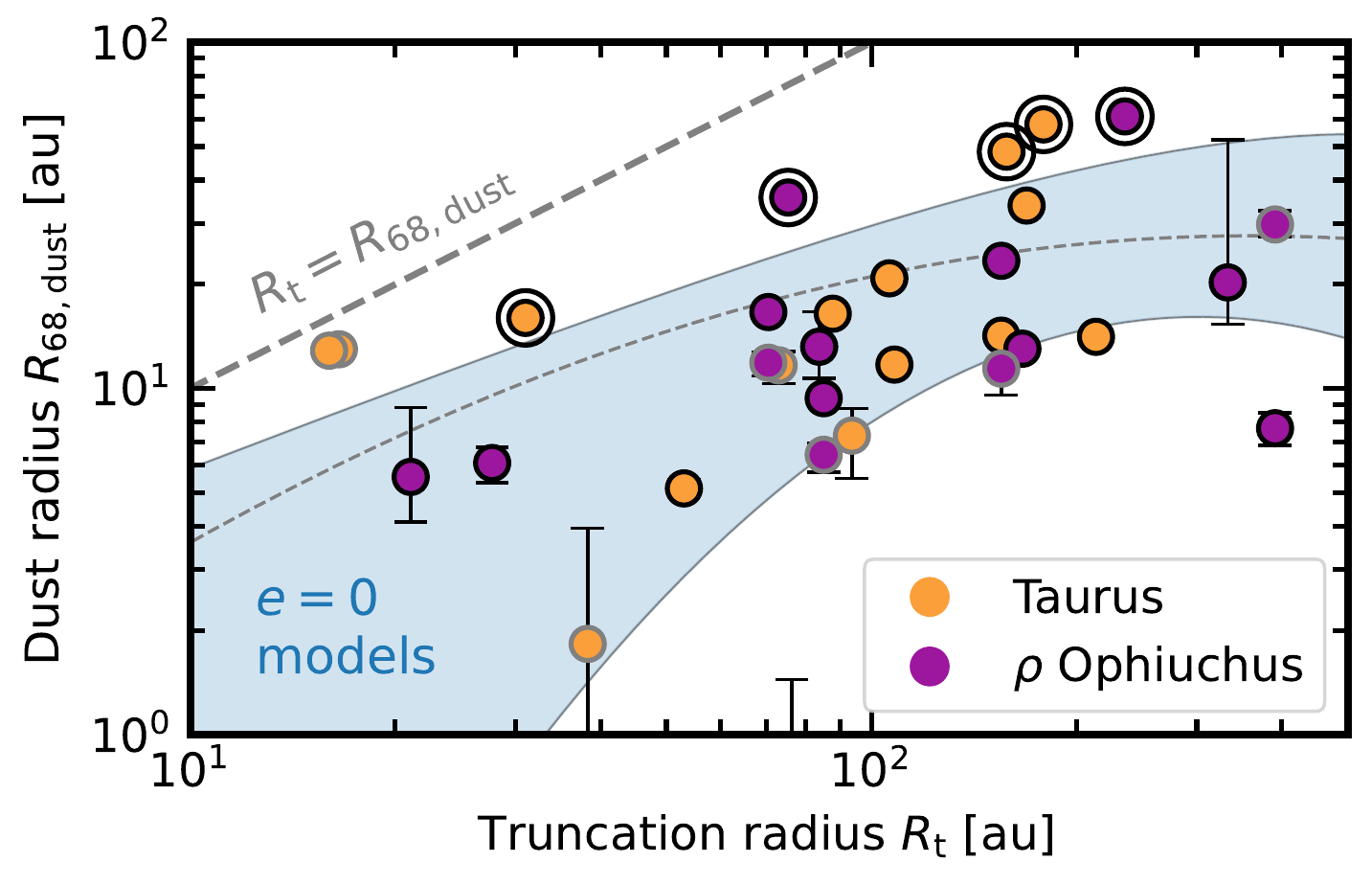}
    \includegraphics[width=0.5\textwidth]{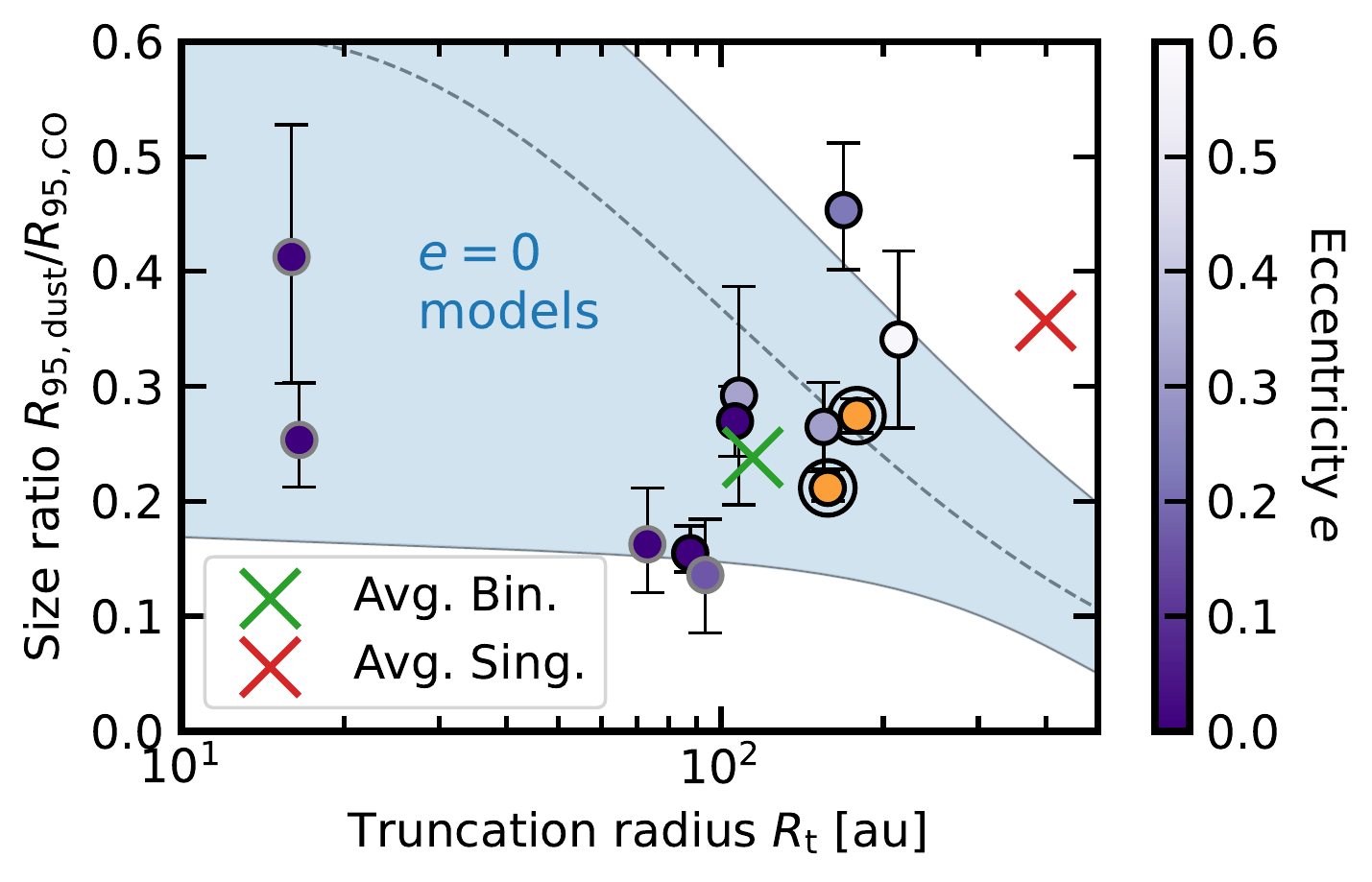}
    \caption{Left: Disc dust sizes in Taurus (orange) and $\rho$~Ophiuchus (purple) binary discs as a function of the truncation radius. The measured sizes are compatible with $e=0$ models (shaded blue region). An outer ring identifies the substructured discs and black/grey marker edges identify primary/secondary discs. Right: Dust-to-gas size ratio as a function of the truncation radius in Taurus. The data points, colour-coded by eccentricity (orange if $e$ is not known), are compatible with $e=0$ models (shaded blue region). Green and red crosses show the average dust-to-gas size ratios in binaries (plotted at the average truncation radius) and singles (plotted at an arbitrary large radius).}
    \label{fig:2}
\end{figure}

These models were post-processed, computing the surface brightness and the dust size (defined as the radius enclosing a faction $x$ of the dust flux, $R_{x,{\rm dust}}$) at $0.85$ mm (ALMA Band 7) by \cite{Zagaria2021_obsv}. In the left panel of Fig.~\ref{fig:2} the median $R_{\rm68,dust}$ and its spread (blue shaded area) are plotted as a function of the truncation radius. Dust disc sizes are smaller than $R_{\rm t}$ (grey dashed line), due to the combined effect of grain growth and radial drift, and are compatible with the observations (orange for Taurus \citealt{Manara2019} and purple for $\rho$~Ophiuchus \citealt{Cox2017}), whose $R_{\rm68,dust}$ is a factor of 2 to 7 \citep{Rota2022} smaller than $R_{\rm t}$. The few outliers can be explained by the presence of gaps or cavities (ringed dots). 

Following-up on previously published tests \citep{Zagaria2021_obsv}, we can use these models to make predictions on the dust-to-gas size ratio. To do so, we define the gas radius as $R_{\rm95,CO}\approx0.97R_{\rm t}$ (for a $T\propto R^{-0.5}$ temperature profile, see Appendix F in \citealt{Trapman2019})\footnote{An accurate estimate of $R_{\rm95,CO}$ requires thermo-chemical simulations and proper post-processing with radiative-transfer codes. Alternatively, one can assume that $^{12}$CO is optically thick through the disc and define the CO radius as the location in the disc where the gas surface density falls below the minimum column density below which $^{12}$CO is not efficiently self-shielded against photodissociation. In our models, this radius is larger than the truncation radius, so we assume it to be $R_{\rm t}$. The prescription of \cite{Trapman2019} is used to convert the total CO radius to $R_{\rm95,CO}$.}. In the right panel of Fig.~\ref{fig:2}, the median size ratio and its spread (blue shaded area) are plotted as a function of the truncation radius. They provide a good agreement with Taurus discs in binaries, colour-coded by their eccentricity, observationally-inferred from CO sizes \citep{Rota2022}. 

This outcome is intriguing provided that in the case of isolated discs substructures need to be invoked to explain the difference between the predicted and observationally inferred dust-to-gas size ratios \citep{Toci2021}. A possible explanation for this behaviour would be that gaps and rings in binary discs are able to survive only in the innermost disc regions. Here the effects of tidal interactions would be milder and allow for substructures to endure \citep{Zsom2011}, as expected from their ability to affect $R_{\rm68,dust}$ (left panel in Fig.~\ref{fig:2}). Instead, in the outer disc, gaps and rings would be disrupted by disc-binary interactions allowing for substantial dust drift, which explains why $R_{\rm95,dust}/R_{\rm95,CO}$ is in agreement with smooth models (right panel in Fig.~\ref{fig:2}). Curiously, if instead $R_{\rm68,dust}/R_{\rm68,CO}$ is considered, single and binary observations provide similar estimates, suggesting that substructures are again important, as expected from $R_{\rm68,CO}$ tracing more internal regions of the disc.

\paragraph{Accretion: explaining the shorter accretion timescale}
The inference of a shorter accretion timescale in binary than in single-star discs \citep{WhiteGhez2001} was recently confirmed with high statistical significance using ALMA and VLT/X-Shooter data in several nearby star-forming regions \citep{Zagaria2022,Gangi2022}. Tidal truncation provides a possible explanation for this behaviour. Shorter accretion timescales are expected in binary discs compared to single ones, because truncation reduces the disc material reservoir without immediately affecting accretion \citep{RosottiClarke2018}. When dust evolution is considered, the accretion time scale is even shortened, because radial drift efficiently removes dust, decreasing the overall dust-to-gas ratio below the standard value used to infer disc masses from dust continuum observations \citep{Zagaria2022}. This effect can be taken into account post-processing the models and measuring their disc masses as is done for data. A comparison shows that, while models explain well the behaviour of Lupus discs (age 1 to 3 Myr), in Upper Sco (age 5 to 10 Myr) the observationally-inferred disc masses are larger than in the models, suggesting that either (undetected) dust traps are present, or that dust is coagulating less efficiently \citep{Zagaria2022}. \\

\noindent All in all, these simple models can reproduce well the behaviour of the binary disc populations in young nearby star-forming regions, even though larger observational samples would be beneficial to get better constraints. Possible limitations are discussed in \autoref{sec:6}.

\section{Circumbinary discs}\label{sec:5}

\subsection{Circumbinary discs in the literature}\label{sec:5.1}

\paragraph*{Circumbinary disc evolution}
``Pure'' circumbinary discs exist only around relatively close binaries ($a\lesssim 10$ to 30~au), as for wider separations the dynamics is significantly altered by the presence of circumstellar discs. Despite their scarcity \citep{Czekala2019}, these systems still represent an important area of study: they are the birthplaces of circumbinary planets, and they serve as a unique laboratory for testing disc physics.

In the case of coplanar circumbinary discs, the main difference from the single-star case is that the binary torque truncates the inner disc, and acts to suppress accretion onto the stars. If no material accretes across the inner cavity, then we have a decretion disc \citep{Pringle1991}, but even relatively low levels of accretion from the inner disc edge lead to significant deviations from a pure decretion disc solution. The efficiency of accretion across the inner cavity is therefore crucial in determining the long-term evolution of circumbinary discs. However, as discussed below, this process is not yet well understood, and is therefore usually parametrised in disc evolution models \cite[e.g.,][]{MacFadyenMilosavljevic2008,Alexander2012}.

Simplified 1D models of circumbinary disc evolution were first presented by \cite{Alexander2012}, and later improved upon by \cite{Ronco2021}. These models incorporated viscous accretion, mass-loss due to photoevaporative disc winds, and the tidal torque from the binary, with accretion at the inner edge parametrised in a relatively simple fashion. The dominant factors in the evolution are the mass-loss rate and the accretion efficiency from the inner disc edge.  High wind rates and/or high accretion efficiencies lead to very short disc lifetimes. By contrast, low wind rates and/or inefficient accretion at the inner edge can result in circumbinary discs living significantly longer than their single-star counterparts.

The existence of long-lived circumbinary discs (at ages up to $\approx10$~Myr) may also place an upper limit on disc mass-loss rates. \cite{Alexander2012} argued that the population of circumbinary discs in Taurus-Auriga sets an upper limit $\dot{M}_{\rm wind} \lesssim M_{\rm disc}/t_{\rm disc} \approx 10^{-9}\ M_{\odot}$~yr$^{-1}$, as higher disc wind rates would result in disc lifetimes much shorter than observed. However, this estimate was based on spatially unresolved observations, and should be revisited with the vastly improved data of the ALMA era. 

\paragraph*{Circumbinary disc dynamics}
The torque exerted by a binary on a disc can efficiently open up a cavity in the circumbinary material, when the binary mass ratio is sufficiently high (precisely for $q>0.04$, \citealt{DOrazio2016}). Interactions with the central binary force circumbinary discs to become eccentric and rigidly precess at slow rates \citep{Thun2017}. Interestingly, the disc eccentricity was found, in turn, to be an important parameter in setting the cavity size \citep{Ragusa2020}. This is reflected in a number of works where the large circumbinary disc cavities are eccentric \citep{Thun2017,Sudarshan2022}. For high-enough mass ratios, high-contrast and long-lived azimuthal overdensities (``lumps'' or ``horseshoes'') moving with Keplerian velocity around the edge of the cavity are formed \citep{Shi2012,Farris2014,Miranda2017,Ragusa2017,Ragusa2020}. These can explain the routinely-observed azimuthal asymmetries in the ALMA data discussed in \autoref{sec:2}. 

Models with circular binaries have the most eccentric cavities. Increasing the binary eccentricity, the cavity becomes less eccentric until a critical value $e\approx0.16$, where the trend reverses and the cavity eccentricity increases again. The eccentricity is expected to be larger in the inner disc and (exponentially) decrease outwards \citep{MunozLithwick2020,Ragusa2020}, as tentatively observed in CS~Cha \citep{Kurtovic2022}. The disc eccentricity is also expected to increase for larger disc to binary mass ratios and to correlate with the cavity size \citep{Ragusa2020}. However, this picture can be altered if planets are present \citep{ThunKley2018,Kley2019}. Planets massive enough to open up a gap shield the outer disc and reduce its eccentricity thus moving closer to the binary in circular orbits (to positions in good agreement with the observed ones \citealt{Penzlin2021}). Lower-mass planets, instead, have eccentric orbits co-precessing with the cavity at large radii.

Most recently, dust evolution in circumbinary discs was also studied. \cite{Chachan2019,Coleman2022} agreed that large dust is efficiently trapped in the pressure bump at the cavity edge \citep{Thun2017}. When similar abundances of dust and gas are considered (assumed to arise because of substantial trapping), the binary cavity shrinks (qualitatively as in the case of self-gravitating discs, \citealt{Mutter2017a,Mutter2017b}) and circularises. This makes it possible for planets to migrate inwards, to the innermost orbit where the three-body binary and planet system is dynamically stable \citep{Coleman2022}.

In misaligned circumbinary discs, dust evolution shows interesting features. Solids with ${\rm St}\gtrsim10$ can be efficiently trapped in ring-like substructures formed by the differential precession of gas and dust \citep{AlyLodato2020}. Analytical and numerical investigations proved that these ``dynamical dust traps'' (i.e., not due to a null pressure gradient) form at two specific locations in the disc, where the projection of dust velocity onto the gas plane equals the gas velocity \citep{Longarini2021}. These predictions are consistent with the recent observations of a misaligned disc in GW~Ori \citep{Kraus2020}. Dust pile-ups are enhanced for higher disc inclinations and eccentricities, with local increase of dust-to-gas ratio up to ten times \citep{Aly2021}.

\paragraph*{Accretion in circumbinary discs} 
The problem of accretion in circumbinary discs has been studied in great detail due to its applicability in different contexts (with scales spanning from supermassive black hole binaries to planet and satellite interactions). In particular, in the case of black hole binaries, during the gas-driven migration phase, the accretion process influences the spin alignment process \citep{Gerosa2015}, determining the gravitational wave frequency pattern and the black-hole recoil after coalescence. For circumstellar discs, how the accretion rate from the circumbinary disc splits between the two stars is especially relevant in the context of binary formation models, since in the early disc phases (Class 0/I) the disc is massive compared to the star, and accretion from the circumbinary disc can potentially change the mass ratio between the two stars. When the mass accreted by the secondary is at least comparable to that accreted by the primary, the system evolves towards mass equalisation.

First analytical studies predicted a suppression of the accretion when high mass ratio companions were present \citep{Pringle1991}. However, these works suffered from the simplified assumption of considering the problem in 1D. Later on, 2/3D simulations, both in the case of stars \citep{ArtymowiczLubow1996,GunterKley2002} and black holes \citep{MacFadyenMilosavljevic2008,Shi2012,DOrazio2013,Farris2014}, showed that streams of circumbinary material can penetrate the disc cavity, allowing for accretion onto the forming binary. These studies suggested that the binary potential does not reduce accretion. However, it was later pointed out that this result depends on the disc aspect ratio. In fact, the accretion rate is progressively suppressed for thinner and thinner discs, while for $H/R>0.1$ it plateaus to the value estimated in the case of single star \citep{Ragusa2016}.

In the latter case, when the stars are at apocenter, and can efficiently interact with the disc, dense clumps of material are launched towards/swept by the binary \citep{MacFadyenMilosavljevic2008,Shi2012,DOrazio2013,Farris2014,MunozLai2016}. Once this material enters the Roche lobe of one of the stars, it can be accreted. However, when a circumstellar disc is present, it acts as a buffer, i.e. smoothing the fast periodicity on the binary orbit timescale in favour of a larger timescale comparable to the cavity edge one \citep{Farris2014,Ragusa2016,YoungClarke2015,MunozLai2016}. As a result, the bulk of material is accreted just before pericenter. This process repeats quasi-periodically, determining a characteristic ``pulsed accretion'' \citep{ArtymowiczLubow1996,MunozLai2016} profile, where apocenter quiescence alternates with accretion bursts about the pericenter.

Several works have argued that accretion takes place preferentially on the secondary star, because it orbits farther from the centre of mass of the binary/closer to the disc \citep{ArtymowiczLubow1996,deValBorro2011,Young2015,YoungClarke2015,Farris2014,Munoz2020}. However, a number of binary and disc parameters can substantially change this picture: higher mass ratios \citep{Farris2014,Duffel2020,Munoz2020} and disc temperatures \citep{Young2015,YoungClarke2015}, or a smaller viscosity \citep{Duffel2020}, lead to comparable accretion rates on each binary component. A similar effect also takes place in non-coplanar systems \citep{Smallwood2022}. In eccentric systems, the precession of the disc cavity makes it possible for either component to accrete the most \citep{Dunhill2015,MunozLai2016}. A symmetry break, i.e., a change in the preferential destination of accretion from one component to the other, is predicted to take place over hundreds of binary orbits in these systems. Eventually a quasi-steady state is established and comparable masses are accreted on the primary and secondary \citep{MunozLai2016}. Finally, accretion results in a net transfer of angular momentum to the binary, that increases its semi-major axis, when $q\gtrsim0.2$ \citep{Munoz2019,Munoz2020}, and circularises eccentric binaries, with the possible exception of moderately eccentric systems ($e\approx0.1$, \citealt{Munoz2019}).

Also in hierarchical triple systems the highest accretor is generally the less-massive component. However, mass equalisation takes place one order of magnitude faster than in binary systems with the same parameters \citep{Ceppi2022}. This is because a close binary generally accretes more than a binary companion of the same mass, as a consequence of the increased geometrical cross section.

In the case of dust, accretion of solids can take place due to the disruption of the dust trap, co-located with the pressure maximum at the cavity edge, by binary interactions. Subdivision of material on each binary component is as in the gas \citep{Chachan2019}. However, this process stops after few binary orbits, once the pressure maximum migrates outwards enough \citep{Thun2017}, on a timescale that increases with binary eccentricity and mass ratio \citep{Chachan2019}.

The predictions of circumbinary accretion simulations discussed so far are difficult to test observationally. While streams of accreting material in multiple stellar systems have been detected \citep{Alves2019,Keppler2020,Reynolds2021,Murillo2022}, signals of pulsed accretion remain extremely rare. At the time of writing only TWA~3A \citep{Tofflemire2017b,Tofflemire2019} and DQ~Tau \citep{Tofflemire2017a,Kospal2018,Muzerolle2019} showed luminosity busts near pericenter that could be attributed to pulsed accretion\footnote{UZ~Tau~E \citep{Jensen2007,Ardila2015} and LRLL~54361 \citep{Muzerolle2013} are currently debated pulsed-accretion candidates.}. Enhancements of their accretion rates up to a factor of ten were measured near pericenter, in agreement with models. However, while in TWA~3A the primary accretes more \citep{Tofflemire2019}, in DQ~Tau the most accreting component, often the secondary, changes on shorter timescales than predicted \citep{Fiorellino2022,Pouilly2022}. This behaviour depends on the emission line used to compute the accretion luminosity, with different lines tracing different stages in the accretion process \citep{Fiorellino2022}. Furthermore, in the case of DQ~Tau accretion events have been occasionally detected at apoastron, as well \citep{Tofflemire2017a,Kospal2018,Pouilly2022}. A possible explanation is its small binary separation. Indeed, in very close binaries the stellar magnetospheres can merge at pericenter \citep{Pouilly2022}, ripping circumstellar discs apart and making accretion more chaotic. According to near infrared observations, also dust is expected to be present in the circumbinary cavity of DQ~Tau, and eventually to be accreted with gas \citep{Kospal2018,Muzerolle2019}. \\

To sum up, the simulations of circumbinary accretion discussed so far depict a complicated process, quasi-periodic but intrinsically variable and dependent on a number of different parameters. Nonetheless, the pulsed accretion model qualitatively agrees with observations. Two main limitations complicate the interpretation of numerical prediction in the case of protoplanetary discs and need to be addressed in the future: (a) very high viscosities are employed (in between those expected in binary black hole and planet-forming discs); (b) magnetic fields are not included. The latter can reduce the size of circumstellar disc, potentially leading to interaction of the accretion streams and the stellar magnetospheres, increasing variability.

\subsection{New models}\label{sec:5.2}
Even though the evolution of a circumbinary disc requires 2/3D simulations to be studied accurately, computationally cheaper methods are necessary to explore the vast parameter space characteristic of this problem. Here we examine if the population-oriented framework recently employed in the case of circumstellar discs \citep{RosottiClarke2018,Zagaria2021_theory} can be adopted also in the circumbinary case. 

We run 1D dusty simulations of viscously-evolving circumbinary discs. The presence of a companion is modelled reducing the accretion rate by 90\% compared to a single-star disc with the same parameters \citep{MacFadyenMilosavljevic2008,Alexander2012}. For dust growth and dynamics, the two-population model \citep{Birnstiel2012} and the Lagrangian advection of the dust fraction \citep{LaibePrice2014,Booth2017} are used. We consider two opposite cases: (a) no limitations on dust accretion, regardless of particle sizes; (b) no accretion of large (cm-sized) grains, as found in recent 2D models \citep{Chachan2019}. 

\begin{figure}[t!]
    \centering
    \includegraphics[width=\textwidth]{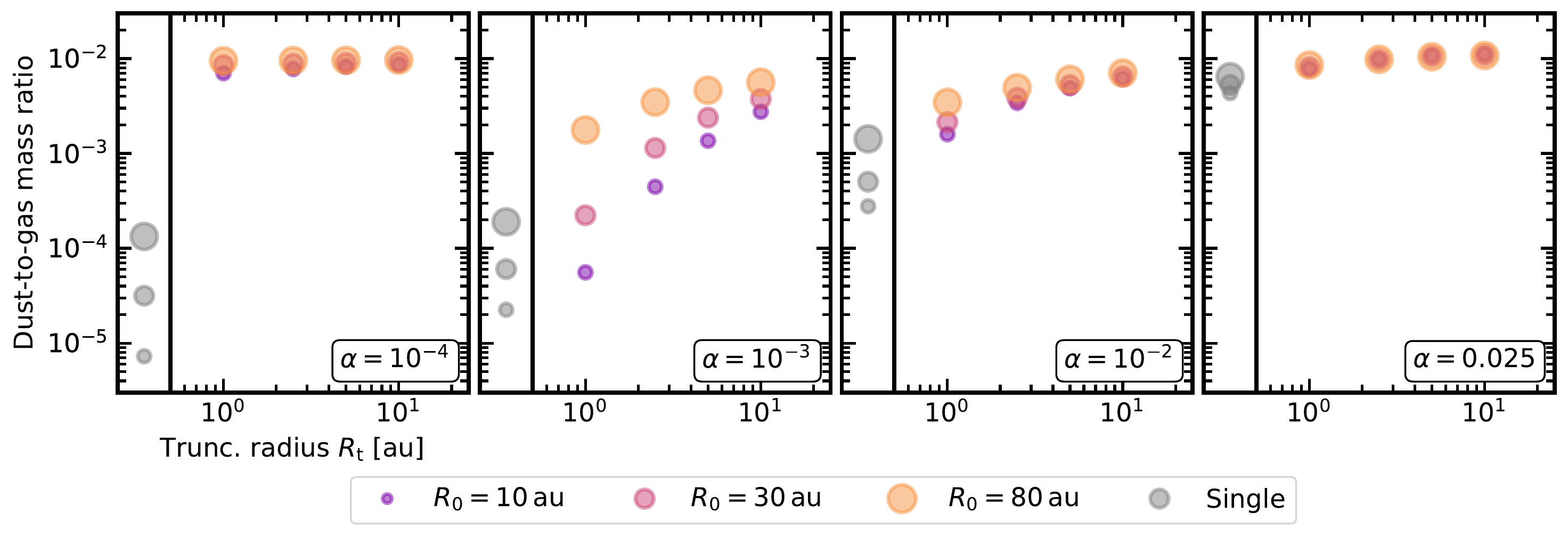}
    \caption{The dust-to-gas mass ratio in circumbinary discs substantially depends on viscosity and only moderately increases with binary separation (2~Myr snapshot). Single discs are in grey while different colours/sizes display different values of $R_0$.}
    \label{fig:3}
\end{figure}

When dust can accrete without restrictions (case a, not shown), after 2~Myr the dust-to-gas mass ratio in circumbinary and single-star discs (as in \citealt{Rosotti2019_sizes}) is the same\footnote{Except for $H/R\geq0.05$, $\alpha\geq10^{-2}$, when the gas evolves differently between the circumbinary and single-star disc case: the inner gas boundary condition effectively produces a \textit{decretion} disc.}. Instead, in the case of limited dust accretion (case b) singles and circumbinary discs behave differently. Our results are shown in Fig.~\ref{fig:3}, where the dust-to-gas mass ratio is plotted as a function of the truncation radius, after 2~Myr, for different values of $\alpha$. The dust-to-gas mass ratio substantially varies with viscosity and only moderately with $R_{\rm t}$.

The dependence on $\alpha$ can be explained in terms of the different, fragmentation- or drift-dominated, dust regime. If the viscosity is small, dust is drift dominated. The closed dust inner boundary favours the accumulation of solids in a narrow ring around $R_{\rm t}$. In this ring, the maximum grain size is set by fragmentation \citep{Pinilla2012}, that replenishes the population of small grains. When $\alpha=10^{-4}$ the gas accretion rate is small and so only few small grains, well coupled with gas, are accreted: the overall dust-to-gas mass ratio is similar to the initial one. Instead, for $\alpha=10^{-3}$, more gas and small grains can be accreted, reducing the overall dust-to-gas mass ratio. If the viscosity is large, dust is fragmentation dominated. An inner ring forms, because of the closed dust inner boundary, but it is wide due to the less efficient drift of dust and the bigger diffusion. So, even though the accretion of gas and small grains is faster, less small grains pile up at the inner disc rim and are accreted. As a consequence, the dust-to-gas mass ratio is larger than for $\alpha=10^{-3}$. Finally, increasing $R_{\rm t}$ reduces the dust loss because of the smaller accretion rate.

Clearly both the case of fully open and fully closed dust inner boundary are extreme, and the behaviour of solids in circumbinary disc is ``in between''. Notwithstanding, we showed that 1D models have the potential to inform us on the available material for planet (\textit{in situ} or \textit{ex situ}) formation in circumbinary discs \textit{on secular timescales}, and how this depends on several disc and binary parameters. However, we need 2/3D-informed dust accretion rates to tailor our simulations to and provide more reliable results.

\section{Discussion}\label{sec:6}

In this section we discuss open questions and future issues.

\paragraph{The observational status: the need for homogeneous surveys for binarity}
It is not an overstatement that, even in the most well studied star forming regions, binarity for the whole young stellar population is still not well known. For example, a comparison of binary fractions among disc bearing stars between Lupus and Upper Sco shows differences (see e.g., appendix A of \citealt{Zagaria2022}) that are best explained by different sensitivities of the binary surveys performed in the region; and even a well studied region such as Chameleon does not have a complete binary survey \citep{Daemgen2016}. Identifying binaries in a more complete and homogeneous way across star forming regions should be a top priority on the observational side. Once such a sample is available, it will be important to assess whether the observational trends and results discussed in this review are confirmed.
Other important issues on the observational side include:
\begin{itemize}
    \item \bfit{Are substructures as common in circumstellar binary discs as they are in single stars?} They certainly have been observed in binary systems \citep{Yamaguchi2021,Rosotti2020,Long2020,Manara2019,Kurtovic2018,Osorio2016}, but their prevalence is still not well known. As discussed in \autoref{sec:4.2} they can be much more internal (at smaller radii) than in single star, which makes their detection more challenging.
    \item \bfit{Why are circumbinary discs rare?} A possible solution would be that some of the observed transition discs are due to a binary inside the cavity. While there is one spectacular example where this is confirmed \citep{Price2018}, it is currently not known how many stellar companions are hidden inside those cavities.
\end{itemize}

\paragraph{The role of MHD winds}

The entirety of the published work we have summarised in the previous sections has considered discs evolving in the viscous framework. As recently reviewed by \cite{ManaraPPVII}, an alternative rapidly gaining traction in the field is instead that discs evolve under the influence of MHD winds. Future work should investigate from the theoretical side what difference this makes to disc evolution in binary systems. While this will require dedicated studies, it is worth reflecting that, in the circumstellar case we highlighted in \autoref{sec:4.2}, the outer boundary condition plays a main role in making binary discs different from those around single stars. The presence of the companion suppresses the viscous expansion of the disc and therefore hastens disc evolution. In the MHD wind scenario there is no viscous spreading and so the evolution of circumstellar discs should be closer to that of single discs, bearing in mind there could be differences in the initial conditions. This could be exploited as an avenue to distinguish between the two mechanisms in the future, if good enough statistics is available on the observational side. In the circumbinary case, there should be smaller differences between the viscosity and MHD winds, but circumbinary discs nevertheless set upper limits on the wind mass-loss rates (see \autoref{sec:5.1}). This constraint comes from studies of photo-evaporating discs \citep{Alexander2012} but should be applicable also to MHD winds, though this needs to be investigated further.

\paragraph{How does grain growth proceed in circumstellar binary discs?}
\cite{Zsom2011} is the only work that has investigated in detail grain growth in circumstellar binary systems, but they were able  to study only short timespans (a few binary orbits). Considering it is now possible to run hydrodynamical simulations with dust grain growth \citep{Drazkowska2019}, an update based on the results of last decade, as well as with simulations run for several dynamical timescales, would certainly be needed in order to assess more thoroughly the impact of binarity on the evolution of disc solids. 

\paragraph{The need for hydrodynamic simulations on secular timescale}
Considering that the presence of a companion clearly breaks azimuthal symmetry, 1D models are forced to take the assumption that there is a well-defined truncation radius at the inner (circumbinary) or outer (circumstellar) boundary, and that accretion through the boundary happens at a prescribed rate (circumbinary) or not at all (circumstellar). As we showed in \autoref{sec:5.2}, the evolutionary of circumbinary discs is sensitive to the details of the accretion rate prescription. This should be informed by hydrodynamic simulations, but a systematic study is still lacking. In the circumstellar case, the assumption that there is no mass transfer between the two discs seems well motivated, but to the best of our knowledge no study has compared the results of disc evolution on secular timescales between the 1D and higher-dimensionality approaches, which can also capture other sources of angular momentum transport such as the spiral arms launched by the companion. The 1D approach also cannot capture that the binary orbits or the cavity \cite[e.g.,][]{ThunKley2018} are often eccentric or inclined with respect to the disc(s) \citep{Jensen2004,AkesonJensen2014,Williams2014,Salyk2014,Takakuwa2017,Kennedy2019,Czekala2019,Rota2022}. Studies in 2D or even 3D proved to be useful guides (e.g., \citealt{Ragusa2020,Hirsh2020} in the case of eccentric discs). However, it is still computationally prohibitive to quantify the impact of these issues on the disc secular evolution. 

\section{Conclusions}\label{sec:7}
We reviewed the main literature results on disc evolution in binary systems with particular focus on the case of dust. 

Observations of circumstellar binary discs agree that the presence of a companion affects their evolution. Discs in binary systems are less-long lived, fainter and smaller than single-star ones, particularly in the case of a close companions with extreme mass ratio. This qualitatively explains the small stellar companion fraction of planet hosts.
\begin{itemize}
    \item Simulations of discs in binary systems agree that in these extreme environment planet formation is difficult. The short disc lifetime and the fast removal of material make challenging to explain planet formation in the context of core accretion and suggest that planetesimals must form fast (see \autoref{sec:4.1}).
    \item Simple population-oriented models can explain well the behaviour of disc fraction with time and the relative primary/secondary disc lifetime as a function of the binary separation and mass ratio \citep{RosottiClarke2018}. When dust is added, disc sizes and accretion timescales can also be reproduced \citep{Zagaria2021_obsv,Zagaria2021_theory,Zagaria2022}.
\end{itemize}

In the case of circumbinary discs the inner binary can extend the disc lifetime, making these system unique laboratories to study planet-formation and the efficiency disc dispersal by thermal winds (see \autoref{sec:5.1}). 

We ran a new a set of simplified 1D models of viscous circumbinary discs inspired by the recent 2D dusty simulations of \cite{Chachan2019,Coleman2022}. We explored how the dust-to-gas mass ratio changes with different initial parameters, showing that it substantially depends on viscosity when the accretion of large grains is inhibited. Better prescription for the dust accretion rate, motivated by 2/3D simulations, are needed to benchmark these models. 

\backmatter

\bmhead{Acknowledgments} We are grateful to the anonymous referee for their comments that helped improving this manuscript. FZ thanks Alessia Rota for sharing binary disc sizes and Enrico Ragusa for his very helpful comments on the draft. FZ is grateful to STFC and Cambridge Trust for his PhD studentship. GR acknowledges support from the Netherlands Organisation for Scientific Research (NWO, program number 016.Veni.192.233) and from an STFC Ernest Rutherford Fellowship (grant number ST/T003855/1). This work was funded by the European Union under the European Union’s Horizon Europe Research \& Innovation Programme grant No. 101039651 (DiscEvol). Views and opinions expressed are however those of the author(s) only and do not necessarily reflect those of the European Union or the European Research Council. Neither the European Union nor the granting authority can be held responsible for them. This work was also supported by the European Union’s Horizon 2020 research and innovation programme under the Marie Sklodowska Curie grant agreement number 823823 (DUSTBUSTERS). \textit{Software}: \texttt{numpy} \citep{numpy20_2020Natur.585..357H}, \texttt{matplotlib} \citep{matplotlib_Hunter:2007}, \texttt{scipy} \citep{scipy_2020SciPy-NMeth}, \texttt{JupyterNotebook} \citep{Jupyter_nootbok}.

\section*{Data Availability Statement}
Data sharing not applicable to this article as no datasets were generated or analysed during the current study.








\bibliography{sn-bibliography}


\end{document}